\documentclass[11pt]{article}
\usepackage{geometry}
\usepackage{graphicx}
\usepackage{color}
\usepackage{multirow}
\usepackage{bm}
\begin{document}
\bibliographystyle{plain}
\title{\Large\bf{A Hybrid Approach for Improved Content-based Image Retrieval using Segmentation}}
\author{Smarajit Bose$^{1}$, Amita Pal $^{1}$\footnote{Corresponding author}, Jhimli Mallick$^{2}$, Sunil Kumar$^{3}$ and Pratyaydipta Rudra$^4$\\
 \\ \footnotesize
$^1$  Applied Statistics Division, Indian Statistical Institute, \\
\footnotesize 203 B. T. Road, Kolkata-700 108, India \\
\footnotesize $^2$  TechBLA Solutions, Kolkata, India\\ 
\footnotesize $^3$ ETH, Zurich, Switzerland\\
\footnotesize $^4$ University of North Carolina, Chapel Hill, USA\\ \\
{\footnotesize E-mail: smarajit@isical.ac.in, pamita@isical.ac.in,}\\
{\footnotesize jhimlimallick.mallick1@gmail.com,}\\ {\footnotesize sunilkr.isi@gmail.com, pratyayr@gmail.com} 
}
\date{}
\maketitle
\thispagestyle{empty}
\begin{abstract}
  {The objective of Content-Based Image Retrieval (CBIR) methods is essentially to extract, from large (image) databases, a specified number of  images similar in visual and semantic content to a so-called \textit{query} image. To bridge the \textit{semantic gap} that exists between the representation of an image by low-level features (namely, colour, shape, texture) and its high-level semantic content as perceived by humans, CBIR systems typically make use of the \textit{relevance feedback} (RF) mechanism. RF  iteratively incorporates user-given inputs regarding the relevance of retrieved images, to improve retrieval efficiency. One approach is to vary the weights of the features dynamically via \textit{feature reweighting}. In this work, an attempt has been made to improve retrieval accuracy by enhancing a CBIR system based on color features alone, through implicit incorporation of shape information obtained through prior segmentation of the images. Novel schemes for feature reweighting as well as for initialization of the relevant set for improved relevance feedback, have also been proposed for boosting performance of RF-based CBIR. At the same time, new measures for evaluation of retrieval accuracy have been suggested, to overcome the limitations of  existing measures in the RF context. Results of extensive experiments have been presented to illustrate the effectiveness of the proposed approaches.}
  
\end{abstract}


\section{Introduction}

As a direct consequence of rapid advances in digital imaging technology, millions of
images are being generated everyday by innumerable sources like defence and civilian satellites,
military reconnaissance and surveillance flights, fingerprinting and facial-image-capturing
devices for security and forensic purposes, scientific experiments, biomedical imaging and home entertainment systems. Large repositories of images have become a commonplace reality due to the availability of cheaper digital storage devices
and the internet. However, maintaining such repositories is meaningless in the absence of methodologies that can enable a user to extract or retrieve information (in the form of images) of interest as and when required.

The first step in this direction was the indexing of image databases using descriptive textual information or \textit{metadata} like captions, keywords, file names and indexing icons~\cite{key-2,key-3}, in a manner similar to cataloging books in a library. The resulting First Generation Visual
Information Retrieval (VIR) systems~\cite{marques2002} were thus text- and concept-based, and the textual
information (metadata) describing an image was used for indexing and searching. This
method, though simple, was subjective and crude at best,  since all the information that a picture or image carries
can not possibly be adequately represented even with "a thousand words". The underlying principle itself was faulty, since images need to be seen and searched as images, in terms of their content. This school of thought led to the advent of the Second Generation VIR systems, or Content-based Image Retrieval (CBIR) systems which, exploit the content to fulfill their objective. These systems support query by content, where the notion of content includes, in increasing order of complexity: perceptual properties (like colour, shape and texture),
semantic primitives (abstractions such as objects, roles, scenes), and subjective attributes
(like impressions, emotions and meaning associated with the perceptual properties)~\cite{marques2002}. The CBIR
system retrieves and presents images similar in some user-defined sense to the query image. The
description of content should serve that goal primarily~\cite{smeulders2000}.

CBIR methods therefore look for images in large databases that are very similar to a supplied query image, where the search is based on the contents of the image rather than metadata. The term \textit{content} in this context might refer to colors, shapes, textures or any other higher-level descriptor(s) that can be derived from the image itself. 

 In a typical CBIR system, features are extracted from each image in the database and stored in the feature database. The same features are extracted from the query image as well. The system computes the distance or the similarity between the feature vectors for the query image and that of each image in the database, and retrieves images (usually a fixed number, specified by the user, known as the \textit{scope} of the system) closest to the query image~\cite{key-13,key-12}.

The low-level features used to represent an image do not necessarily capture adequately the high-level semantics and human perception of that image. This leads to the so-called \textit{semantic gap} in the CBIR context. A solution to this problem is user intervention in the form of \textit{Relevance Feedback} (RF)~\cite{ muller2004,ortega2003,rui1999,yoshitaka1991,zhang2003,zhou2003}. For a given query, the system first retrieves a set of images ranked in order of their similarity to the query image, in terms of  a similarity metric, which represents the distance between the feature vector of the query image and that of each image in the database. Then the user is asked to identify images that are relevant or irrelevant (or non-relevant) to his/her query. The system extracts information from these samples and uses that information to improve retrieval results, and a revised ranked list of images is  presented to the user. This process continues until there is no further improvement in the result or the user is satisfied with the result.  One way of attaining this objective is \textit{feature reweighting}, which essentially assigns greater weights to features that discriminate well between relevant and non-relevant images, thus enhancing retrieval, and smaller weights to those features that do not. Another approach is the instance-based approach, which considers the distance of an image in the database from the query as the minimum distance of the image from the set of all relevant images. This is useful to move through the feature space to the regions with clusters of relevant images. 

        Image features based on a single attribute like color or shape or texture alone are generally not adequate for satisfactory image retrieval~\cite{key-8,key-5,key-15,key-14,key-4}. It has been shown by several researchers in this area that segmentation of the images before matching improves retrieval precision for some image databases. The features derived from each of the clusters obtained by segmentation of the query image are matched with those of the clusters obtained from each image in the database. However, this approach is not uniformly effective for all types of image databases.
				
				This work proposes a modified version of RF-based CBIR with improved retrieval accuracy, through a two-pronged approach. Firstly, a novel approach to feature reweighting for relevance feedback has been proposed, which applies a combination of basic feature reweighting and instance-based cluster density approaches to compute relevance scores and hence weights of features. Secondly, the proposed approach utilizes more of the image content by incorporating both color and shape information. The color information is elicited through the color co-occurrence matrix (CCM) of the image, while the shape information is extracted via segmentation of the image. Three different schemes, based on information obtained from segmentation, are proposed for initialization of the set of retrieved images that is used by RF as a starting point. The efficacy of the proposed approaches has been established through implementation on six different image databases, listed in Table~1. Sample images from some of these are given in Figure~1. This work makes use of the Hue-Saturation-Value (HSV) representation of color images and standard features based on these, which are described in Section~3.4.
				
				Another noteworthy contribution of this work is a couple of new measures for evaluating CBIR methods, which are more appropriate in the context of relevance feedback than the standard measures, \textit{Precision} and \textit{Recall} (Section 2.3).
        
       Organization of the paper is as follows. Section 2 provides an overview of the classical CBIR paradigm based on relevance feedback. Section 3 presents the proposed approaches, together with a couple of new measures of retrieval accuracy. Results are presented in Section 4, while Section 5 summarizes the novelty and effectiveness of the contribution made by this work to CBIR.

\section{Classical Approach to CBIR}
The user of  a typical CBIR system supplies a query image to it  and expects it to extract similar images from a large database. An important component of the system is a feature extraction algorithm which is used to process each image in the database and extract a set of features from it. For an image $I$, let $\bm{f}_I = (f_{I_1},f_{I_2},\cdots, f_{I_d})'$, a $d \times 1$ vector in $I\!\!R^{d}$, be the $d$ features extracted. For a database with $N$ images,  the $d \times N$ matrix $\bm{F}= (\bm{f}_{i1},\bm{f}_{i2},\ldots, \bm{f}_{iN})$, whose $j$-th column is the $d\times 1$ feature vector of the $j$-th image in the database, represents the entire collection of feature vectors that are extracted and stored.  The same feature extraction algorithm is used to process the query image $Q$ too, and the query feature vector is obtained, say, $\bm{f}_{Q} =  (f_{Q_1},f_{Q_2},\cdots, f_{Q_d})'$. The system subsequently uses an appropriate measure to compute the similarity between the query image and each image of the database, and retrieves (a fixed number (specified by the user, known as the \textit{Scope}) images most similar or closest to the query image. 

The inadequacy of the features to represent the perceived content of an image leads to a \textit{semantic gap}, which is bridged through a relevance feedback technique (Section 2.2).

Details of the basic components of a typical CBIR system are discussed briefly in the following sections.

\subsection{Similarity Measure} 
The similarity between the query image $Q$ and any other image $I$ is inversely proportional to the distance between their respective feature vectors. Popular choices of distance measures in CBIR literature are
\begin{eqnarray}
d_1(Q,I) &=& \sum_{j=1}^{d}\,w_{j}\,|f_{Q_j}-f_{I_j}|, \nonumber\\
d_2(Q,I) &=& \sqrt{\sum_{j=1}^{d}\,w_{j}\,(f_{Q_j}-f_{I_j})^2},\label{distancemeasure}
\end{eqnarray}
based on the L1- and L2-norms, respectively. The usual practice is to initialize the weights  as $w_{i} =1/d$. In this work, the distance measure $d_2(Q,I)$ has been used throughout, and has been referred to as $d(Q,I)$ for brevity.

\subsection{Improvement with Relevance Feedback (RF)}

As mentioned earlier, the relatively  low-level features used to represent an image are generally not able to capture adequately its semantic content as perceived by human beings.
This creates the so-called \textit{semantic gap} in the CBIR context. \textit{Relevance Feedback} (RF)~\cite{ muller2004,ortega2003,rui1999,yoshitaka1991,zhang2003,zhou2003} is a commonly-used mechanism which aims to bridge this gap through user intervention. For a given query, the system first retrieves a set of images from the database, ranked in order of their similarity to the query image.  The user is then asked to identify images that are relevant or irrelevant (or non-relevant) to his/her query. The system extracts information from these samples, uses that information to improve retrieval results, and presents a revised ranked list of images to the user. This process is repeated until there is no further improvement in the result or the user is satisfied with the result.   
 
Popular methods for providing this feedback are \textit{feature reweighting} and \textit{instance-based clustering}, which are described below.

\subsubsection{Feature Reweighting}
This widely-used method for implementing relevance feedback assigns different weights to different features~\cite{key-6,key-12}. These weights are modified in each iteration of the relevance feedback. Larger weights are given to those features that discriminate well between relevant and non-relevant images and thus enhance retrieval accuracy. A choice of weights used by Das~\cite{key-6} is based on the ratio of feature variability over all retrieved to the relevant images that are retrieved.    Let ${\sigma_{j}}^{(t)}$ and  ${\sigma_{rel,j}}^{(t)}$, respectively, denote the standard deviations of $f_j$ over the sets  ${\cal R}_t\cup {\cal N}_t$ and  ${\cal R}_t$, where  ${\cal R}_t$ and ${\cal N}_t$ represent the sets of relevant and non-relevant images at the $t$-th RF iteration. A very obvious choice of the weight for the feature $f_j$ at the $(t+1)$-th RF iteration is
\begin{equation}
\label{Wi}
{w_j}^{(t+1)} = \frac{{\sigma_{j}}^{(t)}}{{\sigma_{rel,j}}^{(t)}}.
\end{equation}
When no relevant image (other than the query itself) is retrieved, the denominator is assigned a small positive value $\epsilon$ to avoid the computational problem arising out of  $\sigma_{rel,i}$ becoming zero. The value of $\epsilon$ is chosen such that the weights do not change significantly. 

An efficient way of using both positive and negative samples has been proposed by Wu and Zhang~\cite{key-3}. They used a discriminant ratio to determine the ability of a feature to separate relevant images from the non-relevant ones.   If ${{\cal F}^{(t)}}_{rel,j}=\{f_{I_j},\:I \in {\cal R}_t\}$, the collection of the $j$-th feature of all images in  ${\cal R}_t$, then the dominant range over relevant images at the $t$-th iteration  for the $j^{th}$ feature component is defined as:

\begin{equation}
{D_{j}}^{(t)}=[\min({{\cal F}^{(t)}}_{rel,j}),\max({{\cal F}^{(t)}}_{rel,j})].
\end{equation}

A discriminant ratio (as in \cite{key-12}) can be used to determine the ability of a feature component to separate the relevant images from the non-relevant ones:

\begin{equation}
         {\delta_{j}}^{(t)} = 1 - \frac{\mbox{Number of non-relevant images inside }{D_{j}}^{(t)}}{ |{\cal N}_t|}
         \end{equation}

The value of $\delta_{i}$ lies between 0 and 1.  It is 0 when all non-relevant images are within the dominant range and thus, no weight should be given for that feature component. On the other hand, when there is not a single non-relevant image lying within the dominant range, maximum weight should be given to that feature component.
Based on this, other choices of weights for features are given by
\begin{equation}
\label{W2}
{w_j}^{(t+1)} = \frac{{\delta_{j}}^{(t)}}{\sigma_{{rel,j}}^{(t)}},
\end{equation}
and 
\begin{equation}
\label{W3}
{w_j}^{(t+1)} = {\delta_{j}}^{(t)}*\frac{{\sigma_{j}}^{(t)}}{\sigma_{{rel,j}}^{(t)}}.
\end{equation}

\subsubsection{Instance-Based Methods}
\label{section:instancebasedmethod}

As alternatives to feature reweighting schemes based on Euclidean distances, instance based methods have been quite successful in CBIR, for example, as proposed by Zhang et al.~\cite{zhang2003}. 

Some of the instance-based approaches that have been reported in the literature are described in the next few paragraphs. The purpose of doing so is to lay the groundwork for the proposed combination reweighting scheme developed in Section \ref{proposedreweighting}, which essentially combines the last instance based approach with the reweighting scheme given by~(\ref{W3}) to achieve the highest retrieval performance.

\subsubsection{Minimum Distance from the Set of Relevant Images}
\label{subsection:minimumdistance1}

Here, in each step, the distance of an image in the database from the query is measured by the minimum Euclidean distance of the image from the set of all relevant images. Initially, the set of all relevant images consists of the query image only. This is more useful as compared to using the Euclidean distance from the query image only in the sense that we can move through the feature space to the regions with clusters of relevant images. Thus, if $\cal R$ and $\cal N$ denote respectively the sets of relevant and non-relevant images with respect to the query image $Q$, then
\begin{equation}
\label{distrelset}
d_{R}(Q,I) = \min_{I'\in {\cal R}\cup Q}\,d(I',I)
\end{equation}
where $d(Q,I)$ is as defined in (\ref{distancemeasure}).  

\subsubsection{\textbf{Minimum Distance from the Set of Relevant and Non-Relevant Images}}
\label{subsection:minimumdistance2}
Apart from the minimum distance of an image from the set of relevant images, this method also takes into account the distance from the set of non-relevant images. This is inspired by the observation that the closer an image is to the relevant set and the further it is from the non-relevant set, the more relevant it is. For a database image $I$, if these two distances be $d_{R}(Q,I)$ and $d_{N}(Q,I)$ respectively, then the similarity of the image $I$ with the query image $Q$ is measured by the \textit{relevance score} given by

\begin{equation}
RS(I)=\left({1+\frac{d_{R}(Q,I)}{d_{N}(Q,I)}}\right)^{-1},
\end{equation}
where $d_R(Q,I)$ is as defined in Equation~(\ref{distrelset}), and $d_N(Q,I)$ is the minimum Euclidean distance of $Q$ from the set of non-relevant images, defined as
\begin{equation}
d_{N}(Q,I) = \min_{I'\in {\cal N}}\,d(I',I).
\end{equation}
Clearly the value of the score lies in the interval $[0,1]$. As before, initially the set of relevant images consists of the query image alone, and $d_{N}(Q,I)$ is taken to be 1. The system  retrieves images having maximum relevance scores.

\subsubsection{Instance-based Cluster Density (IBCD) Method}
Even though a small value of $d_{R}(Q,I)$ in the previous method means that image $I$ has a high degree of membership in the relevant set, $d_{R}(Q,I)$ alone may not be able to reflect this completely. For example, an image may be very close to the nearest image of the relevant set, but it may be far away from the centre of the set of relevant images if that nearest image is itself an outlier. That is why it is also desirable that the average distance from all the images in the relevant set is small. Thus the modified relevance score involving cluster density is given by

\begin{equation}
\label{relevancefeedbackcluster}
RS(I)=\left[{1+d_{C}(Q,I) \times \frac{d_{R}(Q,I)}{d_{N}(Q,I)}}\right]^{-1},
\end{equation}
where
$$d_{C}(Q,I)=\frac{1}{|{\cal R}|}\sum_{I'\in {\cal R}} d(I',I).$$
\subsection{Performance evaluation measures}
The two most commonly used measures for evaluating the performance of a CBIR method are \textit{Precision} and \textit{Recall}, which are defined as follows:

\begin{enumerate}
 \item[] $\mbox{Precision}= \displaystyle \frac{\mbox{Number of  relevant images retrieved}}{\mbox{Number of retrieved images}}$
\item[] $\mbox{Recall}=\displaystyle \frac{\mbox{Number of relevant images retrieved}}{\mbox{Total  number  of relevant  images  in the database}}$
\end{enumerate}

Generally the number of images retrieved by any CBIR method (called the  \textit{Scope} of the method) is a prespecified positive integer. Precision and recall values are calculated for each image in the database, and these are averaged over all images (in the database). These averages are conventionally plotted for different values of the scope to provide an illustration of the overall retrieval performance of the method. Usually, the greater the scope,  the larger is the  number of relevant images retrieved, typically leading to increasing values of recall but decreasing values of precision with increasing scope.

However, under relevance feedback, the scenario is slightly different. Here, after the user identifies the relevant and non-relevant images at each iteration, usually a different set (not necessarily disjoint with the earlier set) of images is retrieved in the following iteration  due to change in the search criterion.  This procedure is repeated a number of times after obtaining the relevance feedback from the user after each step. Under this iterative setup, one can still adapt the Precision and Recall measures to have a performance evaluation of the type described above, where these are evaluated for different values of the scope. This can be done by taking the scope at a particular iteration to be the total number of images retrieved till then, and using this scope value for  computing precision and recall on the basis of the total number of relevant images retrieved up to that iteration, for a number of different scope values. The scope value at any given iteration is therefore equal to the number of iterations times $S$,  where $S$ is the initial scope.

There are several issues involved here. For example, it is not desirable to return the same image (relevant or non-relevant) to the user a second time after retrieval at an earlier iteration. Therefore one should aim to retrieve a new set of images at each iteration, which does not contain any of the images retrieved earlier. Further, it makes sense to retrieve only $S - R$ number of images at every step, where $R$ is the current number of relevant images. Under such considerations,  the total number of images to be retrieved changes after every iteration, and it is expected to be different for different images. In view of this, we can expect to see precision-recall plots of the type contained in Figure 4, where typically both increase with iterations (or increasing values of scope), unlike the basic CBIR without RF, where precision decreases but recall increases with increasing scope.  

This discussion clearly establishes that precision and recall are really not appropriate for evaluation of the performance of RF-based CBIR methods, their behaviour becoming counter-intuitive in such cases. Hence we propose two other evaluation measures, defined in the following Section~(\ref{sec_new_measures}), whose behaviour remains consistent, irrespective of whether RF has been used or not.

\section{Proposed Approaches}

\subsection{Segmentation-based similarity}

\subsubsection{Segmentation}
A CBIR system searches image databases for images which have content similar to that of the query image.  So one can expect that its retrieval efficiency to improve after segmentation \cite{key-9,key-8} of the images to identify visually homogeneous sub-regions or objects within them, which are significant indicators of image content. Typically, an image can be segmented in one of two ways, namely, contiguous segmentation and unrestricted segmentation. In contiguous segmentation, adjacent regions that are also similar, are merged recursively to give contiguous homogeneous regions. On the other hand, unrestricted segmentation methods only identify internally homogeneous sub-regions in an image without attempting to merge similar ones. Since these have a lower time-complexity than the contiguous segmentation methods, they have been used in this work. Segmentation typically terminates when objects of interest in an image have been isolated.

\begin{center}
\begin{figure}[h]
\includegraphics[width=6in,height=2in]{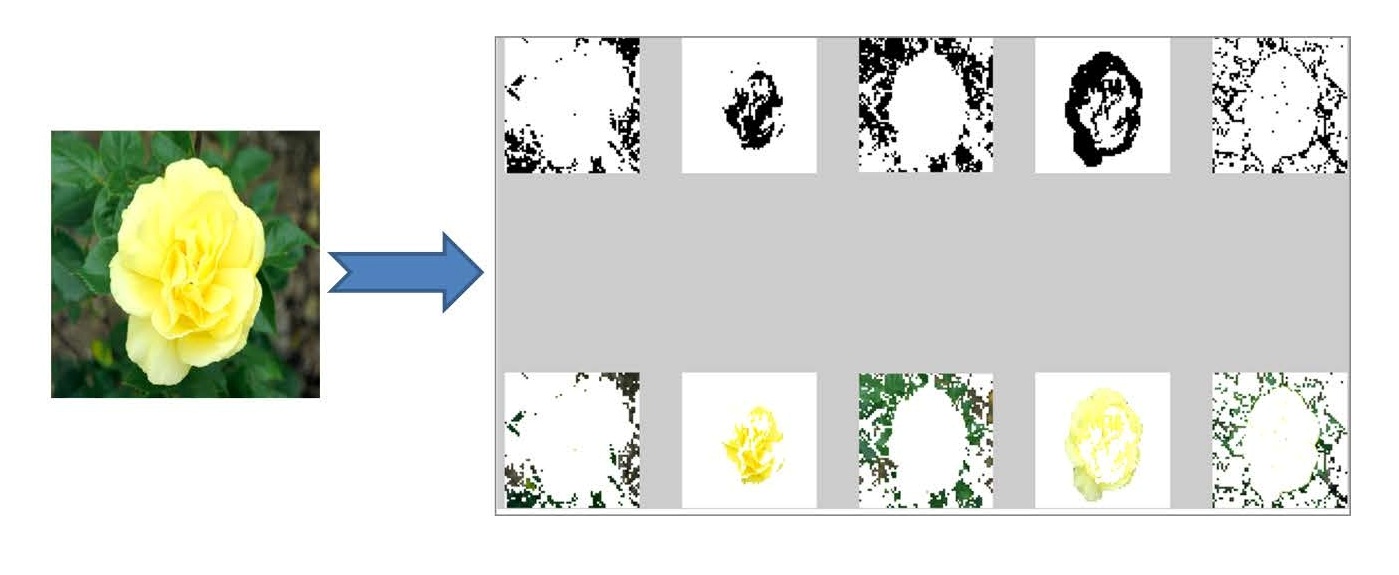}
	\caption{Illustration of  unrestricted segmentation of an image}
\end{figure}
\end{center}

\subsubsection*{Unrestricted Segmentation}
There are many ways to perform unrestricted segmentation on images, like hierarchical clustering, $k$-means clustering and model-based clustering. From the point of view of time-complexity, the  $k$-means algorithm~\cite{key-10} is preferred for segmenting large images. Before clustering, the following preprocessing is done. Each color image of size $N_1\times N_2$ in the HSV color space is split into blocks of size $n_1\times n_2$ (with $n_1=n_2=4$ as suggested in~\cite{key-11}).  A total of $b=N_1N_2/n_1n_2$  blocks is thus obtained. This number is 4096 if $N_1=N_2=256$. The $d$-dimensional feature vector is computed from each block  giving rise to $b$ observations from the image. The  $k$-means algorithm is implemented on this dataset with a prespecified value of $k$, which denotes the number of clusters or segments.  In this work $k$ has been taken to be equal to 8 and empty clusters, if any, are discarded. Figure~1 illustrates how unrestricted segmentation works on a sample image, the images on top showing the distribution of  pixels grouped into the five clusters detected by the algorithm, with the corresponding clusters or image segments shown in the row at the bottom. 

\subsubsection{Proposed Segmentation-based Similarity Measure}
\label{segbasedsimilarity}
The proposed measure of similarity between a query image $Q$ and a database image $I$, based on the distances among their individual segments, is motivated and defined in the following paragraphs.

For the query image $Q$, let $\bm{D}_1(Q,I)$ denote the $n_Q\times n_I$ distance matrix between the $n_Q$ segments of $Q$  and the $n_I$  segments of an image $I$ in the database $\cal D$.

Let $d_{(1)}(Q,I)$ denote the smallest element of $\bm{D}_1(Q,I)$, and let $\rho_{(1)}(Q,I_j),\;j=1,2,\ldots,N$ denote the ranks of $d_{(1)}(Q,I_j),\;j=1,2,\ldots,N$.

Suppose that the minimum $d_{(1)}(Q,I)$ corresponds to the $p_1$-th $Q$-segment and the $q_1$-th $I$-segment, and let $\bm{D}_2(Q,I)$ denote the submatrix obtained by deleting from $\bm{D}_1(Q,I)$ the row corresponding to the $p_1$-th $Q$-segment and the column corresponding to the $q_1$-th $I$-segment. This amounts to removing from further consideration the $Q$-segment and the $I$-segment which are closest to each other.

Likewise, for $i=2,\ldots,r$, $r$ being a prespecified positive integer with $1 \leq r \leq q$, where $q=\min(n_I,\:I \in {\cal D})$, let $d_{(i)}(Q,I)$ denote the smallest element of $\bm{D}_{i}(Q,I)$. Suppose that this minimum corresponds to the $p_i$-th $Q$-segment and the $q_i$-th $I$-segment, and let $\bm{D}_i(Q,I)$ denote the submatrix obtained by deleting from $\bm{D}_{i}(Q,I)$ the row corresponding to the $p_i$-th $Q$-segment and the column corresponding to the $q_i$-th $I$-segment. As before, this amounts to removing from further consideration the $Q$-segment and the $I$-segment which are $i$-th closest to each other.
Let $\rho_{(i)}(Q,I_j),\;j=1,2,\ldots,N$ denote the ranks of $d_{(i)}(Q,I_j),\;j=1,2,\ldots,N$.

 The more similar $Q$ is to $I$, the higher will be the ranks $\rho_{(i)}(Q,I)$, indexed by $i$, for most of the $Q$-segments.


This motivates a new measure of image similarity in the CBIR context, described below.

The proposed segmentation-based distance between the query image $Q$ and an image $I$ in $\cal D$ is defined as
\begin{equation}
d_{seg}(Q,I)= \sum_{i=1}^r \rho_{(i)}(Q,I).
\end{equation}

In this work, $r$ has been taken to be less than or equal to $4$.





The $S$ images $I$ in $\cal D$ with the lowest values of $d_{seg} (Q,I)$ are retrieved in the 1st iteration of relevance feedback in the segmentation-based CBIR approach (referred to as the WS approach) proposed in this work. Retrieval accuracy is expected to increase with increase in the value of $r$ within the range specified above. This is reflected in the outcomes of experiments performed in this work and reported in Section~\ref{results}.

 Henceforth the shorthand notations WOS and WS will be used to denote respectively the conventional CBIR approach (not involving segmentation of images), and the proposed approach based on image segmentation. In both cases, the proposed reweighting scheme (Section 3.1.3) is used for implementing relevance feedback.

\subsubsection{Proposed Feature Reweighting Strategy}
\label{proposedreweighting}
A combination of  the instance-based cluster density (IBCD) method and the reweighting (RW) method is proposed for assigning weights to different features as follows.

Each of the distances $d_R$, $d_N$ and $d_C$  in Equation~(\ref{relevancefeedbackcluster}) is computed as a weighted Euclidean distance  as in Equation~(\ref{distancemeasure})  with weights updated in every iteration by the reweighting scheme given by Equation~(\ref{W3}). The effectiveness of the proposed reweighting scheme (referred to as RW+IBCD for brevity), as compared to simple reweighting (Equation~(\ref{W3})),  is reported in Table~3. The experiments whose results are reported in Tables 4 and 5 also use the proposed RW+IBCD reweighting scheme for relevance feedback.

 It should be noted that the difference between the WOS and WS approaches lies only in the selection of the initial retrieved set for application of RF. The subsequent RF iterations are identical for the two approaches. 

\subsection{Proposed Initialization Schemes for Relevance Feedback}
\label{initialsets}

To exploit additional information on image content, as captured through segmentation, initially WS and WOS methods are applied without RF to retrieve $S$ images each. Based on these two sets of retrieved images, the following alternative methods for specifying the initial set $D_{init}$ (of retrieved images) on which relevance feedback is implemented, are proposed.  All of them lead to improved retrieval accuracy with the proposed WS approach relative to the WOS approach, as will be established empirically in Section~\ref{results}.

 For the WOS approach, seven RF iterations were carried out, the initial retrieval process being treated as the first iteration. RF was applied six times subsequently. However, in two of the initialization schemes proposed below, the number of images retrieved at the beginning is more than the scope $S$, so RF was applied only five times in these cases to ensure a fairer comparison of retrieval accuracies.

\subsubsection{The Intersection  Approach}

Let $D_{WOS}$ and $D_{WS}$ denote sets of $S$ images retrieved by WOS and WS respectively.  Since these two sets are generally quite different, it is expected that the images in the set $D_{inter}=[D_{WOS}\cap D_{WS}]$ have higher chances of being relevant. We select these and some other most similar images from the two sets totaling $S$ for six subsequent RF iterations.  If  $|D_{inter}|=c$, where $|A|$ denotes the cardinality of a set $A$, then $D_{init}$, the initial set of retrieved images presented for RF,  is taken to be equal to
$$D_{init}=D_{inter} \cup {D_{WOS}}^{(1)} \cup {D_{WS}}^{(1)},$$
where ${D_{WS}}^{(1)}$ is the set of  $d_1=[(S-c)/2]$ most similar  images in $D_{WS}-D_{inter}$, and ${D_{WOS}}^{(1)}$ is the set of $d_2=S-c-d_1$ most similar images in $D_{WOS}-D_{inter}$. Here, $[a]$ denotes the largest integer $\leq a$.

 Here, as $|D_{init}|=S$, six iterations of RF are applied.
\subsubsection{The Union Approach}
Here $D_{init}$ is taken to be equal to $D_{union}=D_{WOS}\bigcup D_{WS}$.  Since $S\leq |D_{init}|\leq 2S$, only five iterations of RF are implemented. 

\subsubsection{The Combination Approach}

In this approach, both $D_{WS}$ and $D_{WOS}$ are presented separately for RF, accounting for the first  two iterations. If the number of relevant images in $D_{WS}$ is greater than or equal to that of $D_{WOS}$, feature reweighting is performed with the sets of relevant and non-relevant images in $D_{WS}$ only. Otherwise, they are taken from $D_{WOS}$. 

 Here, $|D_{init}|=2S$ and hence only five iterations of RF are carried out.
\subsection{Proposed Performance Evaluation Measures}
     \label{sec_new_measures}
Motivated by the discussion in the preceding section, the following new measures are proposed for assessing the accuracy of retrieval of a CBIR system:

\begin{enumerate}
	\item $\mbox{Retrieval Efficiency (RE)}= \displaystyle \frac{\mbox{Number of relevant  images  retrieved}}{\mbox{Scope}}$
	\item $\mbox{False Discovery (FD)}= \displaystyle \frac{\mbox{Number  of  non-relevant  images retrieved}}{\mbox{Total number  of  retrieved images}}$
\end{enumerate}
 
Retrieval Efficiency is expected to increase with the number of RF iterations and should converge fast in a few iterations if RF is effective.
 
False discovery, being the ratio of the number of non-relevant images retrieved to the total number of retrieved images, is a measure of erroneous retrieval (that is, the retrieval of non-relevant images), and should be as small as possible.
\subsection{Features Used}
Feature selection is an extremely crucial aspect of CBIR. In this work standard features used in CBIR, as described below, have been adopted.

As expected, features like colour, shape and texture are key indicators of content. An important representation of the spatial distribution of colour in an image is provided by the colour co-occurrence matrix  (CCM)~\cite{key-2,key-4,key-3}. The  $L \times L$ CCM of an image having $L$ colour levels in any one of the dimensions of the $HSV$ (Hue, Saturation, Value) colour space, denoted by $P=[p_{ij}]$, is such that $p_{ij}$ represents the proportion of  pixels with colour level $i$ co-occurring with other pixels with colour level $j$, at a relative position, say, $d$.  The diagonal elements of the CCM give the colour distribution in the image, while the non-diagonal elements convey shape information, since colour changes between adjacent pixels indicates the possible existence of an object edge.  The feature vector used consists of 
all $L$ diagonal elements of the CCM as well as a single number to represent the information contained in its non-diagonal elements, defined as
\begin{equation}
ave\_ndiag=\sum^{L-1}_{i=1}\sum^{L}_{j=i+1}(i+j)p_{ij},
\end{equation}
where $i$ and $j$ are row and column indices.

It has been observed by researchers that $L_H=16$ and  $L_S= L_V=3$  are good choices for number of quantization  levels of $H$, $S$ and $V$ for specifying co-occurrence matrices. A co-occurrence distance  $d=1$ has been used in this work and pixel pairs in both vertical and horizontal directions have been considered, leading to symmetric co-occurrence matrices. Thus only upper diagonal elements  of the CCMs needed to be considered.

Consequently,  $D=(16+1+3+1+3+1)=25$ features were used in this work, following~\cite{key-6}. 

\subsection{Image Databases Used}

To demonstrate the effectiveness of the proposed approach, a number of databases were used, which are listed and briefly described in Table I.

\begin{table}[h]
	\centering
	\caption{List of Image Databases used}
	\label{ListOfImageDatabasesUsed}
\vspace{2mm}
		\begin{tabular}{|l|c|c|c|l|} \hline
		Name & Size & No. of  & Size per & Remarks \\ 
		&& Categories & Category &  \\ \hline
	  DB2000 & 2000 & 10 & 200 & \\
    DB2020 &  2020 & 12 & 96--376 & \\
    DBCaltech & 8365 & 93 & 26--871 & \\
    DB3057 &  3057 & 14 & $\geq$99  & Subset of DBCaltech \\
    DB5276 & 5276 & 33 & 80--798 & Images from several databases$^1$ \\
 DB3767 & 3767 & 17 & 85--798 & Subset of DB5276 \\ \hline
		\end{tabular}
		
		$^1$\scriptsize{DBCaltech, Dinosaur database (containing 99 images of dinosaurs), DB2000 and DB2020}
	
\end{table}

Figure~2 gives illustrative instances of images three of these databases.  There are two images per category for the databases DB2000 and DB2020 whereas for DBCaltech, a single image from each category in a subset of size 25 out of its 93 categories, is shown, just to give an idea of the diversity in each database.
 
\begin{figure}
\centering
\includegraphics[width=10mm,height=10mm]{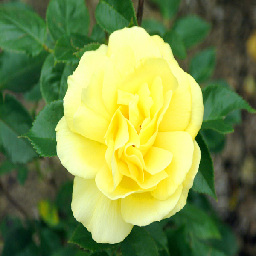}
\includegraphics[width=10mm,height=10mm]{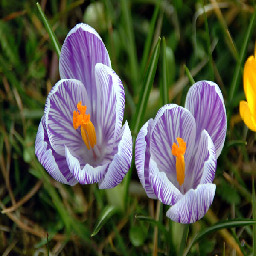}
\includegraphics[width=10mm,height=10mm]{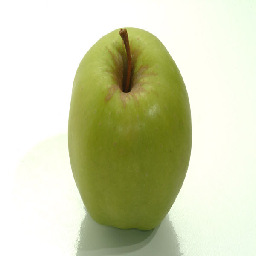}
\includegraphics[width=10mm,height=10mm]{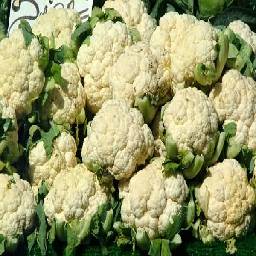}
\includegraphics[width=10mm,height=10mm]{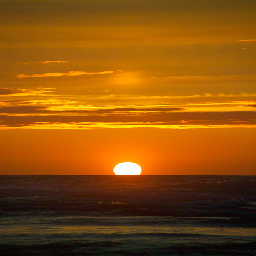}\\
\includegraphics[width=10mm,height=10mm]{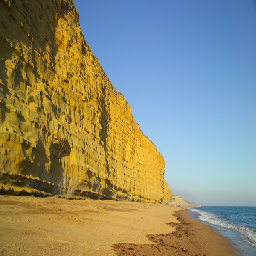}
\includegraphics[width=10mm,height=10mm]{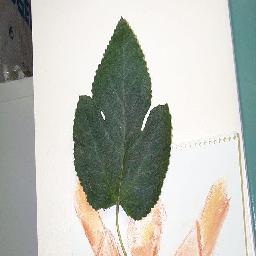}
\includegraphics[width=10mm,height=10mm]{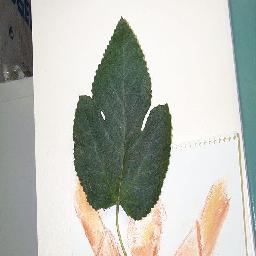}
\includegraphics[width=10mm,height=10mm]{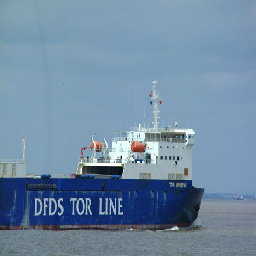}
\includegraphics[width=10mm,height=10mm]{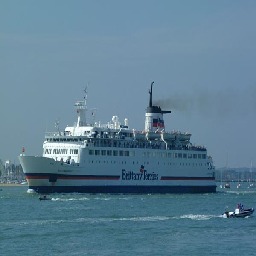}\\
\includegraphics[width=10mm,height=10mm]{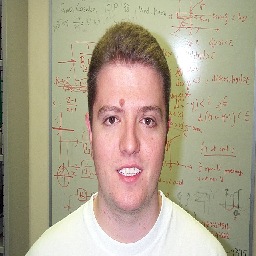}
\includegraphics[width=10mm,height=10mm]{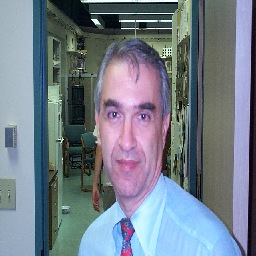}
\includegraphics[width=10mm,height=10mm]{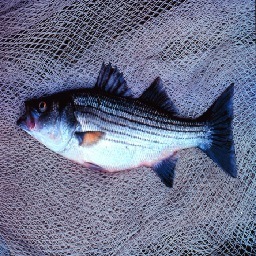}
\includegraphics[width=10mm,height=10mm]{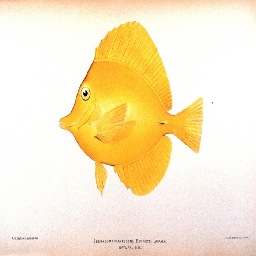}
\includegraphics[width=10mm,height=10mm]{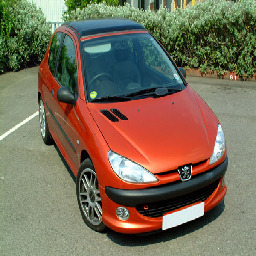}\\
\includegraphics[width=10mm,height=10mm]{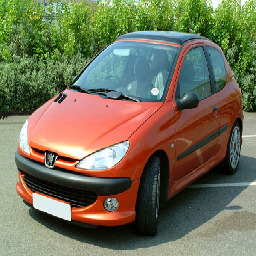}
\includegraphics[width=10mm,height=10mm]{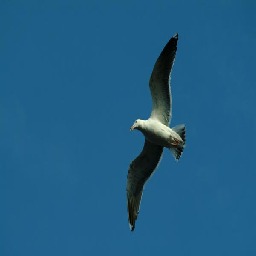}
\includegraphics[width=10mm,height=10mm]{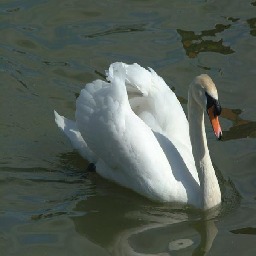}
\includegraphics[width=10mm,height=10mm]{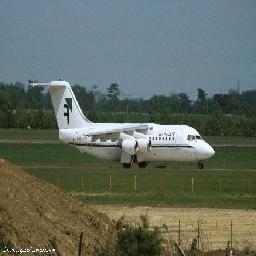}
\includegraphics[width=10mm,height=10mm]{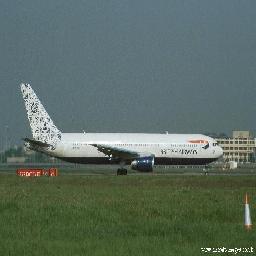}
\begin{center}
(a)
\end{center}
\vspace*{.10in}
\includegraphics[width=10mm,height=10mm]{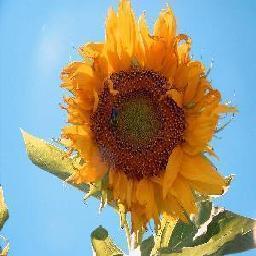}
\includegraphics[width=10mm,height=10mm]{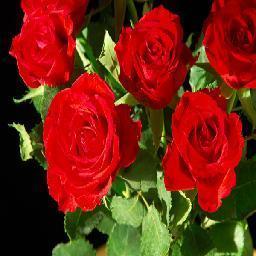}
\includegraphics[width=10mm,height=10mm]{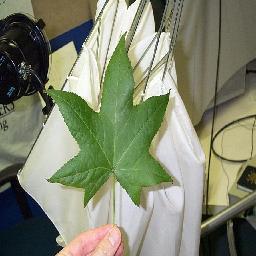}
\includegraphics[width=10mm,height=10mm]{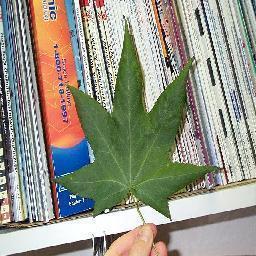}
\includegraphics[width=10mm,height=10mm]{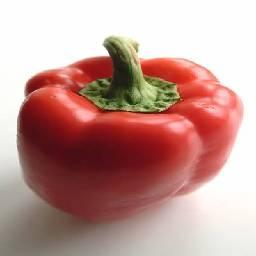}\\
\includegraphics[width=10mm,height=10mm]{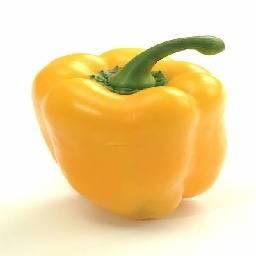}
\includegraphics[width=10mm,height=10mm]{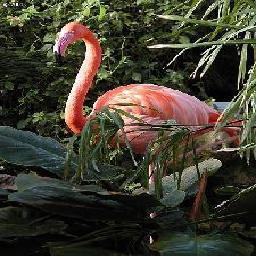}
\includegraphics[width=10mm,height=10mm]{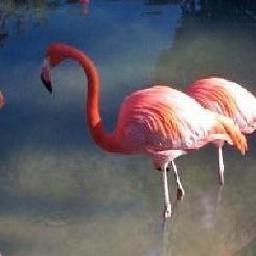}
\includegraphics[width=10mm,height=10mm]{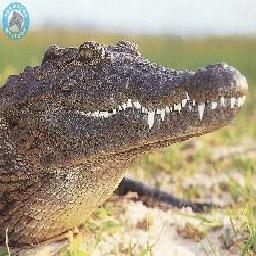}
\includegraphics[width=10mm,height=10mm]{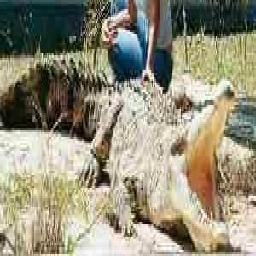}\\
\includegraphics[width=10mm,height=10mm]{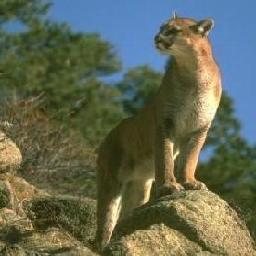}
\includegraphics[width=10mm,height=10mm]{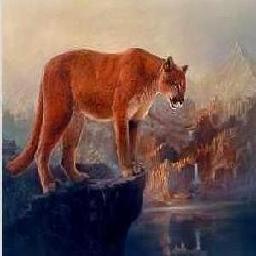}
\includegraphics[width=10mm,height=10mm]{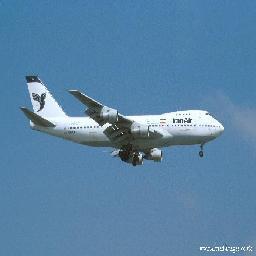}
\includegraphics[width=10mm,height=10mm]{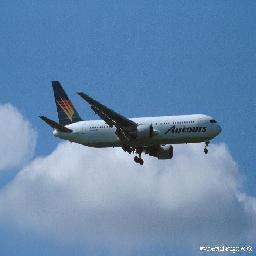}
\includegraphics[width=10mm,height=10mm]{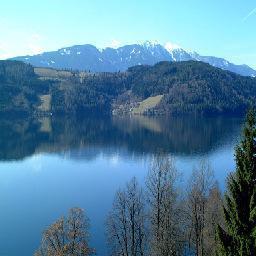}\\
\includegraphics[width=10mm,height=10mm]{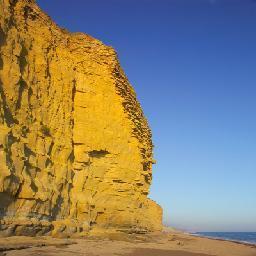}
\includegraphics[width=10mm,height=10mm]{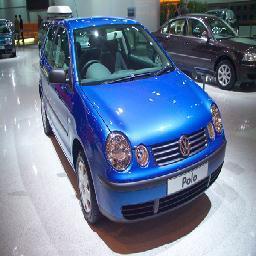}
\includegraphics[width=10mm,height=10mm]{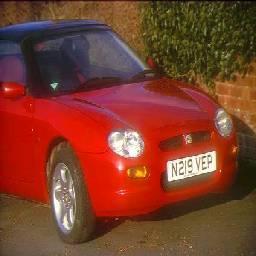}
\includegraphics[width=10mm,height=10mm]{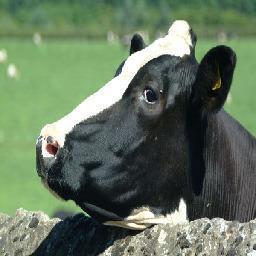}
\includegraphics[width=10mm,height=10mm]{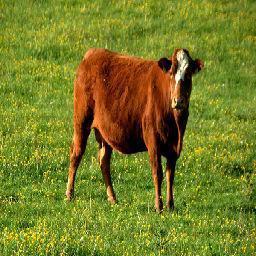}\\
\includegraphics[width=10mm,height=10mm]{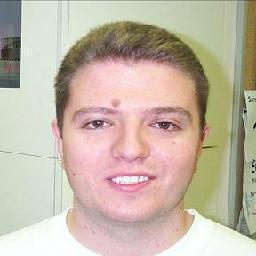}
\includegraphics[width=10mm,height=10mm]{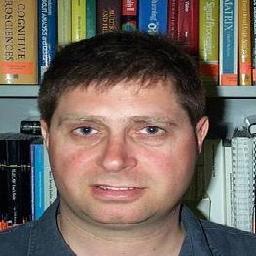}
\includegraphics[width=10mm,height=10mm]{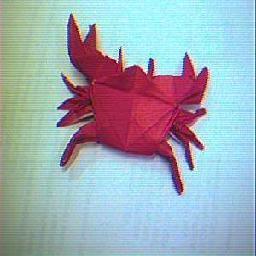}
\includegraphics[width=10mm,height=10mm]{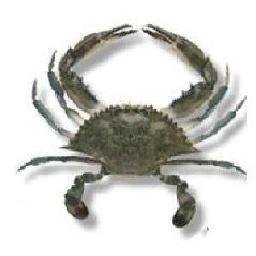}
\includegraphics[width=10mm,height=10mm]{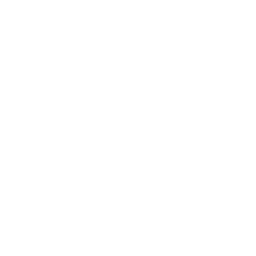}
\vspace*{.10in}
\begin{center}
(b)
\end{center}
\includegraphics[width=10mm,height=10mm]{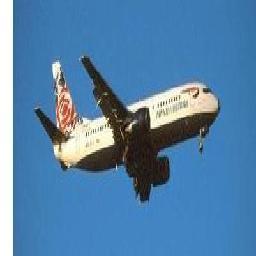}
\includegraphics[width=10mm,height=10mm]{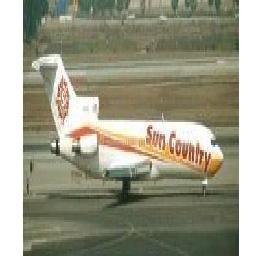}
\includegraphics[width=10mm,height=10mm]{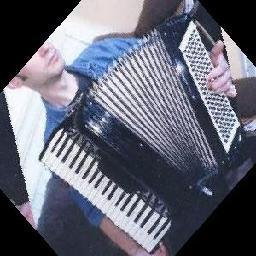}
\includegraphics[width=10mm,height=10mm]{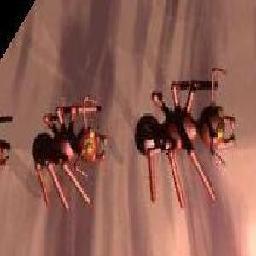}
\includegraphics[width=10mm,height=10mm]{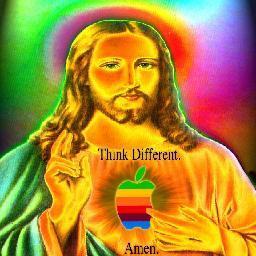}\\
\includegraphics[width=10mm,height=10mm]{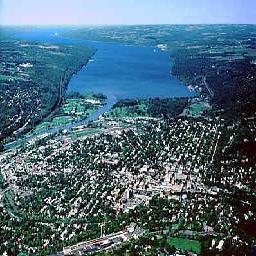}
\includegraphics[width=10mm,height=10mm]{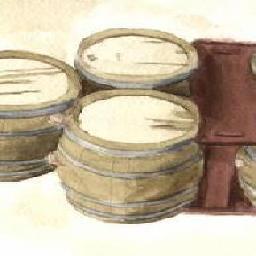}
\includegraphics[width=10mm,height=10mm]{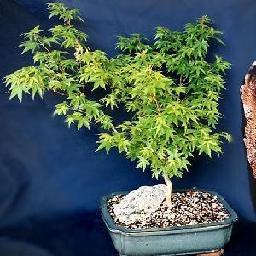}
\includegraphics[width=10mm,height=10mm]{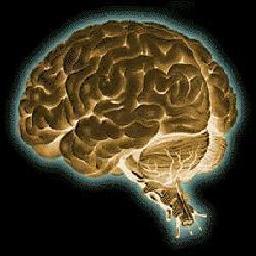}
\includegraphics[width=10mm,height=10mm]{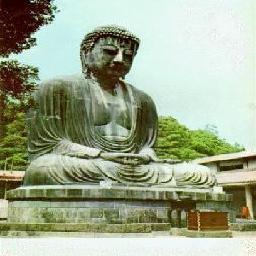}\\
\includegraphics[width=10mm,height=10mm]{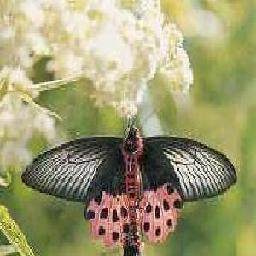}
\includegraphics[width=10mm,height=10mm]{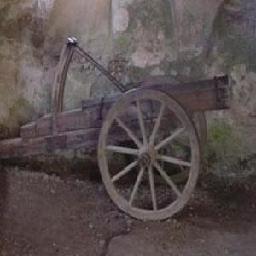}
\includegraphics[width=10mm,height=10mm]{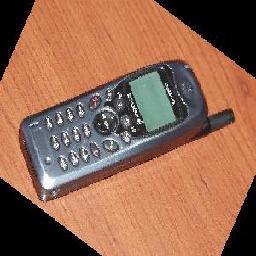}
\includegraphics[width=10mm,height=10mm]{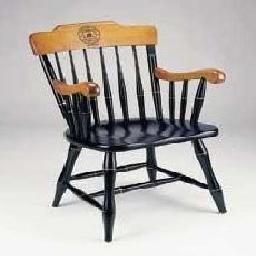}
\includegraphics[width=10mm,height=10mm]{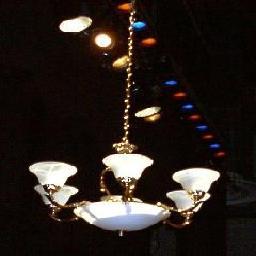}\\
\includegraphics[width=10mm,height=10mm]{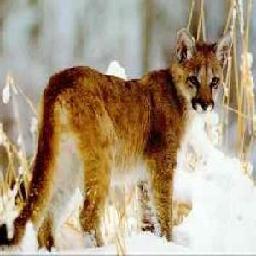}
\includegraphics[width=10mm,height=10mm]{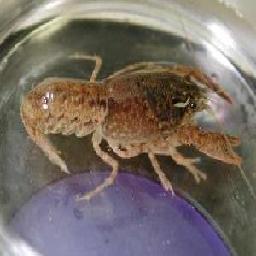}
\includegraphics[width=10mm,height=10mm]{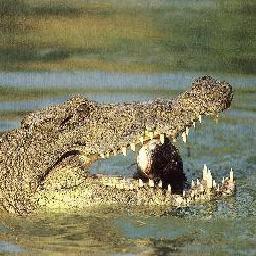}
\includegraphics[width=10mm,height=10mm]{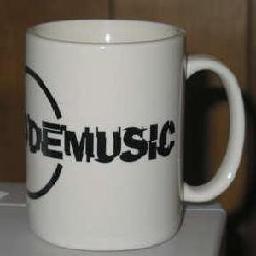}
\includegraphics[width=10mm,height=10mm]{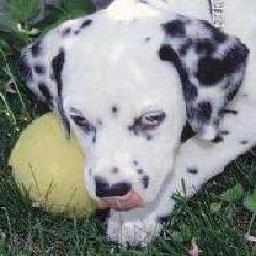}\\
\includegraphics[width=10mm,height=10mm]{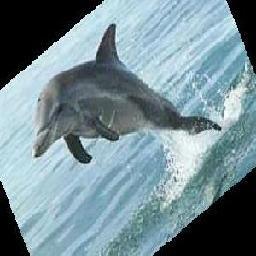}
\includegraphics[width=10mm,height=10mm]{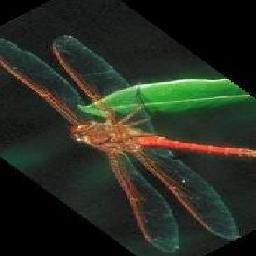}
\includegraphics[width=10mm,height=10mm]{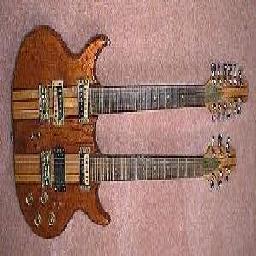}
\includegraphics[width=10mm,height=10mm]{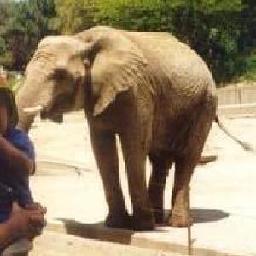}
\includegraphics[width=10mm,height=10mm]{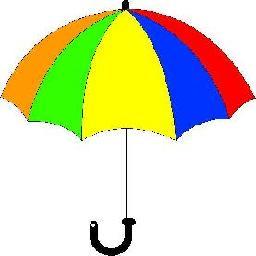}
\begin{center}
(c)
\end{center}
\caption{Sample images from (a) DB2000 (b) DB2020 and (c) DBCaltech.}
\end{figure}   
        
\section{Results}
\label{results}
Table~\ref{EffectOfronRE} shows how an increase in $r$ improves retrieval accuracy, thereby providing empirical justification for the statements made in Section~\ref{segbasedsimilarity}.
\begin{table}[h]
	\centering
	\caption{Effect of $r$ on Relative Efficiency}
	\label{EffectOfronRE}
	\vspace{2mm}
	
		\begin{tabular}{ |l | p{10mm} | p{10mm} | p{10mm} | p{10mm} |} \hline
 \multirow{3}{*}{Database} & \multicolumn{4}{|c|}{Retrieval  Efficiency} \\
& \multicolumn{4}{|c|}{(after 1 iteration) with } \\
\cline{2-5}
& {$r$=1} & {$r$=2} & {$r$=3} & {$r$=4} \\
\hline
     DB2000  &  36.38 &  44.08 &  48.47 &  51.06\\
   \hline
    DB2020 &  31.26 &   38.30 &   42.23 &  44.73\\
 \hline
     DBCaltech &   13.27 &   17.89 &   20.37 &    21.82\\
     \hline
    DB3057 &   33.19 &  38.73 &    41.76 &    43.40\\
   \hline
  DB5276 &     21.32 &      26.56 &     29.84 &    31.69\\
  \hline
 DB3767 &     32.81 &    40.03 &   44.24 &    46.51\\
 \hline
		\end{tabular}
\end{table}

The effectiveness of the proposed reweighting scheme for RF (described in Section~\ref{proposedreweighting}), as compared to simple reweighting in the context of the WOS approach, is reported in Table~3, which contains results obtained after 7 RF iterations.

\begin{table}[h]
\label{table3}
	\centering
	\caption{Effectiveness of the Proposed Reweighting Scheme for RF}
	\vspace{2mm}
	\small
		\begin{tabular}{ | l | c | c | c | c |}
\hline
    \multirow{3}{*}{Database} & \multicolumn{2}{|c|}{Retrieval Efficiency} & \multicolumn{2}{|c|}{False Discovery} \\
    \cline{2-5}
        &  {with simple} & {with proposed} &  {with simple} & {with proposed} \\
        & reweighting & reweighting & reweighting & reweighting \\
				& (RW) & (RW+IBCD) & (RW) & (RW+IBCD)\\
\hline
DB2000 
 & 89.20
 & 94.69
 & 48.61
 & 41.89 \\
\hline
DB2020 
 & 80.32
 & 86.42
 & 56.67
 & 50.93 \\
\hline
DBCaltech 
 & 39.40
 & 42.72
 & 79.02
 & 76.16 \\
\hline
DB3057 
 & 74.65
 & 80.03
 & 54.97
 & 51.08 \\
\hline
DB5276 
 & 62.41
 & 68.26
 & 67.46
 & 62.77 \\
\hline
DB3767 
 & 83.86
 & 90.06
 & 51.08
 & 45.10 \\
\hline		
		\end{tabular}
\end{table}

\begin{figure}[h]
	\centering
		\includegraphics[width=4in,height=3in]{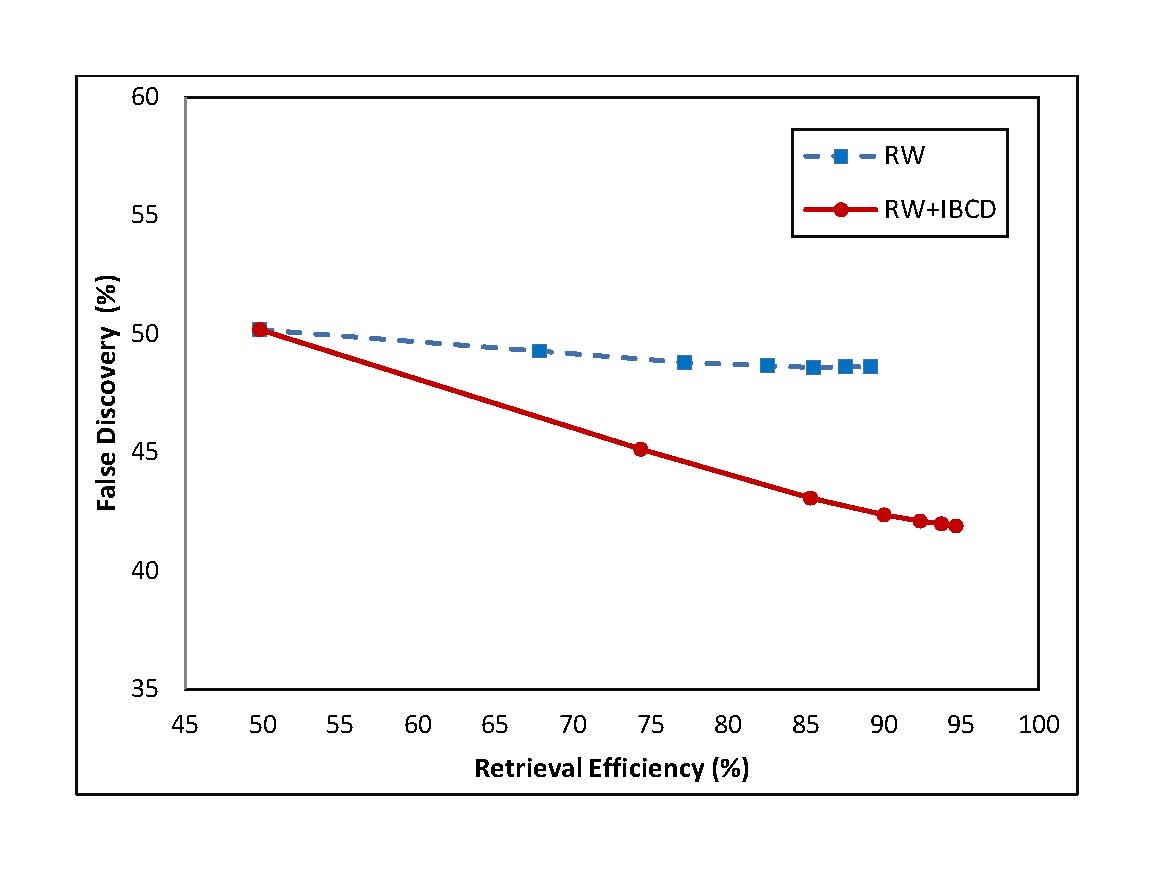}
	\caption{Effectiveness of the Proposed Reweighting Approach}
	\label{fig3}
\end{figure}

These results are presented graphically in Figure~\ref{fig3} for the DB2000 database. It is amply evident from the table as well as the figure that the proposed reweighting scheme performs much better than the basic reweighting. There is a more marked change in  the gain in RE and the drop in FD with every iteration when the proposed reweighting (RW+IBCD) is used.

A comparison between the conventional WOS approach and the proposed WS approach to CBIR, using  the three initialization schemes proposed in Section~\ref{initialsets}, is reported in Tables~\ref{finalresults1} and~\ref{finalresults2}. The feature reweighting scheme described in Section~\ref{proposedreweighting} is used for both. In these tables, the shorthand names $\mbox{WS}_{inter}$, $\mbox{WS}_{union}$ and $\mbox{WS}_{comb}$ are used to identify the WS method initialized by the \textit{intersection}, \textit{union} and \textit{combination} approaches, respectively.  Improvement in retrieval accuracy with the proposed approach is evident in all cases. With respect to False Discovery, we note that it is higher for $\mbox{WS}_{union}$ as compared to WOS, though the former shows better performance in respect of Retrieval Efficiency. However, it is encouraging to note that both $\mbox{WS}_{inter}$ and $\mbox{WS}_{comb}$ are successful in simultaneously reducing False Discovery and achieving higher Retrieval Efficiency relative to WOS. Of these two,   $\mbox{WS}_{comb}$ is clearly performing the best in both respects. As far as Precision and Recall are concerned, again both $\mbox{WS}_{inter}$ and $\mbox{WS}_{comb}$ clearly outperform WOS. However, $\mbox{WS}_{union}$ lost out marginally on Precision while achieving better Recall than WOS. With respect to these measures too  $\mbox{WS}_{comb}$ is found to perform the best.

\begin{figure}[h]
\centering
\includegraphics[width=5in,height=3in]{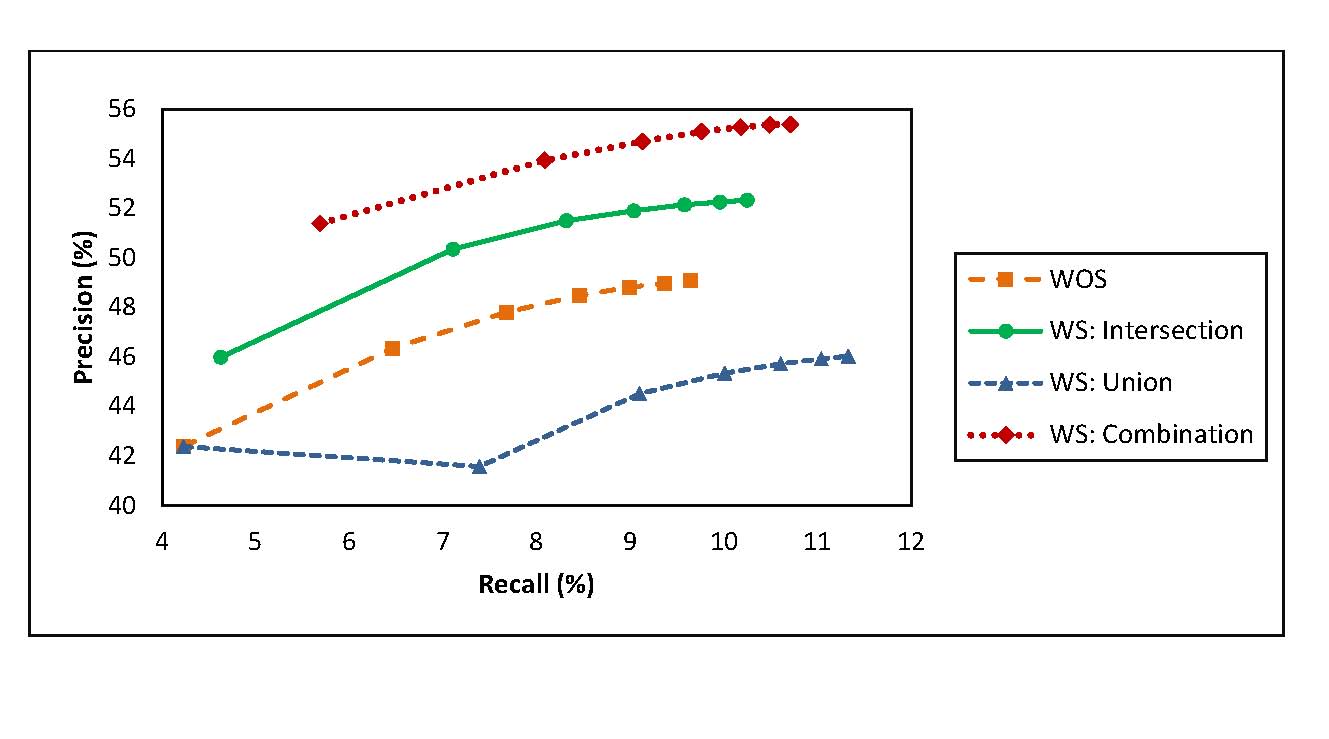}
	\caption{Accuracy with Proposed Approach for DB2020}
\label{fig4}
\end{figure}

\begin{table}[h]
	\centering
	\caption{Effectiveness of the Proposed Segmentation-based Approaches in terms of Proposed Measures}
	\vspace{2mm}
	\label{finalresults1}
	\small
		\begin{tabular}{ | l | c | c | c | c | c | c | c | c |}
\hline
    \multirow{2}{*}{Database} & \multicolumn{4}{|c|}{\textit{Retrieval Efficiency} with} & \multicolumn{4}{|c|}{\textit{False Discovery} with} \\
    \cline{2-9}
        &  {WOS} & $\mbox{WS}_{inter}$ & $\mbox{WS}_{union}$ & $\mbox{WS}_{comb}$ &  {WOS} & $\mbox{WS}_{inter}$ & $\mbox{WS}_{union}$ & $\mbox{WS}_{comb}$\\
\hline						
DB2000 	 & 94.69	 & 96.1	 & 96.58	 & 97.26	 & 41.89	 & 39.62	 & 46.89	 & 35.71 \\
\hline								
DB2020 	 & 86.42	 & 90.34	 & 91.61	 & 91.94	 & 50.93	 & 47.67	 & 53.97	 & 44.61 \\
\hline								
DBCaltech 	 & 42.72	 & 43.98	 & 45.93	 & 46.55	 & 76.16	 & 75.21	 & 77.77	 & 73.33 \\
\hline								
DB3057 	 & 80.03	 & 81.41	 & 83.02	 & 84.49	 & 51.08	 & 50.28	 & 54.04	 & 46.67 \\
\hline								
DB5276 	 & 68.26	 & 69.73	 & 71.84	 & 72.43	 & 62.77	 & 61.66	 & 65.52	 & 58.91 \\
\hline								
DB3767 	 & 90.06	 & 91.03	 & 92.13	 & 92.78	 & 45.1	 & 44.07	 & 49.63	 & 40.67 \\ \hline
		\end{tabular}
\end{table}

The iteration-wise results for the image database DB2020  are presented graphically in Figure~\ref{fig4} to illustrate the typical trends observed in improvement with the proposed approach, in terms of precision and recall. Incidentally, the precision and recall values for the last (7th) iteration are given in the second row of Table~\ref{finalresults2}.

 \begin{table}[hb]
	\centering
	\caption{Effectiveness of the Proposed Segmentation-based Approaches in Terms of Conventional Measures}
	\vspace{2mm}
	\label{finalresults2}
	\small
		\begin{tabular}{ | l | c | c | c | c | c | c | c | c |}
\hline
    \multirow{2}{*}{Database} & \multicolumn{4}{|c|}{\textit{Precision} with} & \multicolumn{4}{|c|}{\textit{Recall} with} \\
    \cline{2-9}
        &  {WOS} & $\mbox{WS}_{inter}$ & $\mbox{WS}_{union}$ & $\mbox{WS}_{comb}$ &  {WOS} & $\mbox{WS}_{inter}$ & $\mbox{WS}_{union}$ & $\mbox{WS}_{comb}$\\
\hline						
DB2000 	 & 58.11	 & 60.38	 & 53.11	 & 64.29	 & 9.47	 & 9.61	 & 10.76	 & 9.75 \\
\hline								
DB2020 	 & 49.07	 & 52.33	 & 46.03	 & 55.39	 & 9.64	 & 10.25	 & 11.33	 & 10.71 \\
\hline								
DBCaltech 	 & 23.84	 & 24.79	 & 22.23	 & 26.67	 & 3.95	 & 4.13	 & 4.77	 & 9.33 \\
\hline								
DB3057 	 & 48.92	 & 49.72	 & 45.96	 & 53.33	 & 6.17	 & 6.35	 & 7.27	 & 7.47 \\
\hline								
DB5276 	 & 37.23	 & 38.34	 & 34.48	 & 41.09	 & 6.56	 & 6.79	 & 7.59	 & 8.72 \\
\hline								
DB3767 	 & 54.9	 & 55.93	 & 50.37	 & 59.33	 & 7.68	 & 7.8	 & 8.7	 & 8.16 \\
\hline								
    \end{tabular}
 \end{table}

 \section{Conclusions}
There are different approaches for Content Based Image retrieval available in the literature. Given a query image, the conventional methods extract features from the entire images, and retrieve those images from the database which are most similar to the query image. Methods based on segmentation of the images have also been proposed where both the query and the images in the database are first segmented, and then the segments thus obtained from the images in the database are matched with segments obtained from the query image. 

In this work, a new hybrid approach for CBIR is proposed in which the conventional approach has been combined with a segmentation-based approach. A relevance feedback mechanism based on feature reweighting with an instance-based distance is employed. Several schemes for combining the two approaches are proposed, and their effectiveness is illustrated with a variety of databases. The proposed approach was successful in improving the retrieval accuracy significantly.

\section*{Acknowledgments}

The authors would like to put on record their indebtedness to Prof. Siddheswar Ray of the Clayton School of Information Technology
at Monash University, Melbourne, Australia, for invaluable discussions, and to his former Ph.D. student, Gita Das, for crucial insights into the CBIR problem via her Ph.D. thesis. The contributions of Dr. Sarat Dass of the Michigan State University,  and Sayantan Banerjee of the North Carolina State University, are also gratefully acknowledged.

\bibliography{cbirrefs}
  \end{document}